\documentclass[useAMS,usenatbib]{mn2e}

\usepackage[dvips]{graphicx}
\usepackage{wrapfig}

\title[Formation of Globular Clusters]
{Formation of globular clusters induced by external ultraviolet radiation}
\author[Kenji Hasegawa, Masayuki Umemura and Tetsu Kitayama]
{Kenji Hasegawa $^{1}$\thanks
{E-mail: hasegawa@ccs.tsukuba.ac.jp (KH); 
umemura@ccs.tsukuba.ac.jp (MU); kitayama@ph.sci.toho-u.ac.jp (TK)}, Masayuki Umemura$^{1}$
and Tetsu Kitayama$^{2}$
\footnotemark[1]\\
$^{1}$Centre for Computational Sciences, University of Tsukuba, Ten-nodai, 1-1-1 Tsukuba, 
Ibaraki 305-8577, Japan\\
$^{2}$Department of Physics, Toho University, Funabashi, Chiba 274-8510, Japan}
\begin{document}

\date{Accepted 2009 May 7. Received 2009 May 1; in original form 2008 July 3}

\pagerange{\pageref{firstpage}--\pageref{lastpage}} \pubyear{2009}

\maketitle

\label{firstpage}

\begin{abstract}
We present a novel scenario for globular cluster (GC) formation, where
the ultraviolet (UV) background radiation effectively works so as to 
produce compact star clusters.
Recent observations on the age distributions of GCs indicate 
that many GCs formed even after the cosmic reionization epoch. 
This implies that a significant fraction of GCs formed in UV background
radiation fields. Also, the star formation in
an early-generation of subgalactic objects may be affected by strong UV radiation 
from preformed massive stars, e.g., Population III stars. Here, we explore
the formation of GCs in UV radiation fields. For this purpose,
we calculate baryon and dark matter (DM) dynamics in 
spherical symmetry, incorporating the self-shielding effects by solving
the radiative transfer of UV radiation. 
In addition, we prescribe the star formation in cooled gas components 
and pursue the dynamics of formed stars. 
As a result, 
we find that the evolution of subgalactic objects in UV background
radiation are separated into three types, that is, 
(1) {\it prompt star formation}, where less massive clouds
($\sim 10^{5-8} M_\odot$) are promptly self-shielded and undergo star formation,
(2) {\it delayed star formation}, where photoionized massive clouds 
($\ga 10^{8} M_\odot$) collapse despite high thermal pressure 
and are eventually self-shielded to form stars in a delayed fashion, 
and 
(3) {\it supersonic infall}, where photoionized less massive clouds 
($\sim 10^{5-8} M_\odot$) contract with supersonic infall velocity and
are self-shielded when a compact core forms. 
In particular, the type (3) is a novel type found in the present
simulations, and eventually produces a very compact star cluster.
The resultant mass-to-light ratios, half-mass radii, and velocity dispersions 
for the three types are compared to the observations of GCs, 
dwarf spheroidals (dSphs), and ultra-compact dwarfs (UCDs). 
It turns out that the properties of star clusters resulting from
{\it supersonic infall} match well with those of observed GCs,
whereas the other two types are distinct from GCs.
Hence, we conclude that {\it supersonic infall} in a UV background 
is a promising mechanism to form GCs. 

\end{abstract}


\section[]{Introduction}
Globular clusters (GCs) are characterized by their compactness,
old age, and low metallicity. Although the mass range of GCs 
($10^{5-6}M_\odot$) is similar to that of dwarf spheroidal galaxies (dSphs),
GCs have much higher stellar velocity dispersions than those of dSphs. 
In addition, the velocity dispersions in GCs 
correlate with their luminosities in a definitely different fashion from
that in dSphs. 
Furthermore, the survey of Fornax galaxy cluster has revealed a 
new class of dwarf galaxies, that is, ultra compact dwarf galaxies (UCDs), 
which are more compact than normal dwarf galaxies but clearly distinct 
from GCs \citep{Drinkwater03}. 
According to the virial theorem, the velocity dispersion is 
expressed as $\sigma_* \propto \sqrt{GM/r_{\rm h}}$, where $M$ and 
$r_{\rm h}$ are the total mass and the half-mass radius of a system. 
The velocity dispersion is then related to the luminosity 
by $\sigma_* \propto L^{1/3}$, assuming $M \propto r_{\rm h}^3$ and 
$L\propto M$. Observed GCs, however, show the correlation like 
$\sigma_* \propto L^{1/2}$ \citep{Djo97,MvdM05}.  
This correlation can be derived if GC radii are almost independent of mass 
\citep{AZ01}.
Interestingly, such a correlation is found not only in Galactic GCs but also in 
GCs in external galaxies \citep{Hasegan05}.  Also, the $\sigma_*-L$ 
relation of UCDs obviously deviates from the virial theorem 
\citep{Drinkwater03}.  As for dSphs, some follow $\sigma_* \propto 
L^{1/3}$, but the majority of dSphs show higher velocity 
dispersions than expected with the virial theorem 
\citep{Mateo98,Matin06,SG07,MCI06,Majewski07}.  These imply that 
GCs as well as UCDs and many dSphs did not form through simple 
gravitational collapse and virialization. 

Various scenarios for the formation GCs have been hitherto considered by 
many authors
\citep[e.g.,][]{PD68,FR85, AZ92, ML92, BC02, KG05,Saitoh06}. 
They are mainly divided into three categories, 
based on the formation epoch of GCs relative to that of a host galaxy.  
The first one is a merger scenario, in which GCs form in the gas-rich mergers of disk 
galaxies \citep{AZ92}. 
In this scenario, interstellar gas is highly compressed by shocks induced 
by the mergers, and very compact star clusters are expected to form. 
Such events likely occur at a later stage of galaxy formation, 
therefore formed star clusters would be young. 
This scenario can explain young star clusters observed in merging or 
starbursting galaxies. However, it might be difficult to explain old 
GC populations. 
Moreover, the metallicity of star clusters is expected to be
the same as that of merging galaxies, which may be
higher than the GC metallicity of $Z \simeq 10^{-4}-10^{-2}Z_\odot$. 

In the second scenario, GCs form at the early stage of 
the host galaxy evolution. 
\cite{FR85} firstly have shown the possibility of GCs formation 
caused by thermal instability, in which proto-globular clusters 
are formed from dense gas clouds with $T\sim 10^4{\rm K}$ 
confined by diffuse hot gas with $T\sim 10^6{\rm K}$. 
In this case, the Jeans mass for such dense clouds roughly corresponds to 
the typical mass scale for observed GCs ($10^{5-6}M_{\odot}$). 
In this model, the clouds should be metal-poor 
to maintain the cloud temperature as $T\sim 10^4{\rm K}$. 
A similar idea have been explored by \cite{ML92}, in which 
the temperature of dense clouds is maintained as $T\sim 10^4{\rm K}$ by photoheating. 
\cite{KG05} have carried out high resolution simulations, and shown that 
GCs can form in high-density cores of giant molecular clouds in dense gaseous 
disk of galaxies. 
They also have found that the resultant ratio of the total GC mass to the baryonic 
mass of the host galaxy is as small as $\approx 10^{-4}$, 
as indicated by observations \citep{Mc99}. 

The last one is the pregalactic formation scenario that has been widely explored, 
in which GCs form in dark matter (DM) minihaloes before they infall into galactic haloes. 
This is a plausible scenario for the formation of low-mass star clusters, and
it can account for many properties of GCs such as low metallicity and old age 
comparable to the cosmic age \citep[e.g.,][]{BC02, MS05,Saitoh06}. 
However, proto-GCs in this scenario are expected to be dominated by dark matter (DM), and 
possess the mass-to-light ratios of $M/L\sim 10$.
Therefore, it does not provide a direct explanation for observed GCs that have 
mass-to-light ratios as low as $M/L\sim 1$ \citep{PM93}. 
Recent simulations of galaxy formation by \cite{Saitoh06} have indicated  
that diffuse DM haloes are selectively stripped by tidal interactions with their host galaxy, 
and baryon-dominated star clusters are naturally explained.  
However, the spatial resolution of the simulations  cannot allow us 
to discuss the internal structure of proto-globular clouds. 
Therefore, the physical reason for condensations of the proto-globular clouds 
is still uncertain. 
The stripping of DM haloes also has been studied using $N$-body 
calculations by \citet{MS05}, where
a hybrid cluster composed of stars and DM is simulated
in the external tidal field of the host galaxy. 
They assumed initially a compact baryon core embedded in a diffuse DM halo,
and explored the dynamical evolution using $N$-body simulations. 
They have found that the surface brightness profiles 
and mass-to-light ratios for the evolved hybrid GCs are consistent with those of observed GCs. 
However, it has not been clarified yet how such compact stellar systems 
form at first. 
Since the virial radii of collapsed objects at $z\sim10$ are 
several hundred parsecs in the mass range of $10^{5-6}M_\odot$, 
that are $\sim10-100$ times larger than the sizes of observed GCs. 
Hence it is not easy to form baryon-dominated star clusters. 
An additional physical mechanism seems to be requisite
to produce a compact stellar core. 
Moreover, for all the above scenarios, 
the physical mechanism to explain the $\sigma_*-L$ 
relation of GCs has not been elucidated.

The ages of Galactic GCs show a broad distribution with the mean age of 
about 12.3Gyr \citep{Puzia05}. A significant fraction of GCs seem
to form after the cosmic reionization epoch $z_{\rm r} \simeq 11$, which is
inferred by {\it Wilkinson Microwave Anisotropic Probe} (WMAP) 
three year data \citep{Page07} and five year data \citep{Komatsu09}. 
Therefore, not a few GCs are thought to form in UV background radiation fields. 
Also, the formation of GCs itself might contribute much to 
the reionization of the universe at redshift $z\approx 6$ \citep{Ricotti02}. 
Furthermore, even before the reionization, the star formation in
an early-generation of galaxies can be affected by strong UV radiation from
preformed massive stars including Population III (Pop III) stars. 
Thus, it is of great significance to consider the formation of GCs in UV radiation
fields.
When a gas cloud is photo-ionized, 
the cloud is heated up to $\sim 10^4$K \citep{MI84, TW96}. 
In addition, soft UV radiation (Lyman-Werner band) destructs $\rm H_2$ molecules, 
which is the main coolant below $10^4$K in low metallicity gas
\citep{Stecher67,Haiman97,Kitayama01}. 
Then, the formation of low mass objects with the virial temperature 
less than $10^4$K is suppressed. 
Moreover, no star is expected to form in ionized gas. 
Hence, to form star clusters, the gas cloud should be self-shielded 
from UV background radiation \citep{TU98}. 
To treat the self-shielding, we should solve the radiative transfer properly.
\citet{Kitayama01} explored the formation of dwarf galaxies within UV radiation fields, 
using spherical symmetric hydrodynamics coupled with the
radiative transfer. With the radiation hydrodynamic (RHD) simulations,  
they showed that dwarf galaxies can form even after the reionization, 
if the self-shielding effectively works. 
\citet{SU04} also studied this issue by three-dimensional RHD simulations including 
formation of stars and the stellar dynamics.  
In their simulations, they showed that a dwarf galaxy with the stellar mass
of $\approx 10^6M_\odot$ is able to form even in UV radiation.
The half-mass radius of the stellar system is $\sim100$pc, 
which is corresponding to the sizes of dSphs, but
greater by about one-order magnitude than the observed GC size. 
Hence, a further mechanism is required to produce much more
compact star clusters in UV radiation.
%Susa \& Kitayama (2000) estimated the core size which can be cooled 
%by $H_2$, by solving the radiative transfer under the assumption of 
%isothermal collapse of a gas cloud. 
%They conclude that strong UV radiation is needed to form GC size cooled core.   
%However, they did not consider the influence of dark matter (DM), 
%which is generally expected to be included in the progenitor of GCs 
%in the case of pre-galactic scenario. 
%In addition, the relaxation of the star cluster was not explored.  
%Therefore, it has not been cleared yet whether compact star cluster such as a GC forms 
%within UV radiation fields.

In this paper, we explore a possibility that a supersonically contracting cloud 
produces a compact star cluster in UV radiation. 
A gas cloud with its infall velocity exceeding the sound speed of ionized gas 
($T\simeq10^4$K) can keep contracting, even if the gas cloud is fully ionized. 
The contracting cloud is eventually shielded from both ionizing photons 
and $\rm H_2$ dissociating photons, and then the cloud can cool via 
$\rm H_2$ cooling, generating a star cluster. 
The star cluster formed through such a process becomes 
very compact due to strong energy dissipation. 

This paper is organized as follows.  In \S 2, the simulation code and 
numerical procedure are described. 
The results of simulations are presented in \S 3, and it is shown
that the evolution of subgalactic objects are typically
separated into three branches. 
The comparison with the observations of GCs, dSphs, and UCDs
is given in \S 4.
The specific frequencies of GCs, required UV intensity,  
tidal effects by host galaxy, and internal feedbacks are discussed in \S 5. 
\S 6 is devoted to the conclusions. 
Throughout this paper, we assume a $\Lambda$-CDM cosmology with 
the matter density $\Omega_{\rm M} = 0.3$, the cosmological constant 
$\Omega_{\lambda}=0.7$, the Hubble constant $h=0.7$, 
and the baryon density $\Omega_{\rm B}=0.05$.

\section[]{Simulations}
\subsection{Numerical Scheme}
We use a radiation-hydrodynamic scheme developed by \cite{Kitayama01},
to solve hydrodynamics coupled with radiative transfer. 
This scheme is based on the second-order Lagrangian finite-difference method in 
spherical symmetry. In this scheme, we treat self-consistently 
gravitational force of dark matter and baryon, hydrodynamics, 
non-equilibrium chemistry of primodial 
gas including $\rm H_2$, and the radiative transfer of ionizing photons. 
For self-shielding by $\rm H_2$ against photo-dissociating photons, 
we employ the self-shielding function introduced by \cite{DB96}. 
[See also \cite{Kitayama01} for the details of scheme.]
Initially, the number of shells is $N_{\rm b}=600$ for gas component 
and $N_{\rm d}$=10000 for DM component. 
Shocks are treated with artificial viscosity. 
In order to investigate whether the results are sensitive to 
the treatment of artificial viscosity, we perform the simulations with 
stronger artificial viscosity (10 times as strong as the fiducial one) or 
weaker artificial viscosity (0.1 times as weak as the fiducial one). 
As a result, we find that the changes of results are only a few percent. 
Thus, we conclude that the present results are not sensitive to the 
treatment of artificial viscosity.

We also consider the formation of stars from cooled gas component and  
pursue the relaxation process of formed stars. 
We take the following conditions as the star formation criteria:
\begin{equation}
	T_{\rm g}<2000K,
\end{equation} 
\begin{equation}
	v_{\rm r}<0, 
\end{equation}
\begin{equation}
	\frac{d\rho_{\rm g}}{dt}>0, 
\end{equation}
\begin{equation}
	\frac{dT_{\rm g}}{dt}<0.
\end{equation}
Here $T_{\rm g}$, $v_{\rm r}$, and $\rho_{\rm g}$ are the temperature, 
the infall velocity, and the density of gas shell. 
Ionized gas cannot cool below $10^4$K, 
if satisfying the condition (3). 
Hence, the gas must be shielded against UV radiation to satisfy the condition (1). 
In that sense, the condition (1) is a most important criterion. 
However, the threshold temperature $2000$K is not so crucial. 
When we changed this threshold temperature from $1000$K to $5000$K, 
we found no significant difference in the main results. 
We assume that if a gas shell satisfies all of the above conditions, 
the gas shell becomes a stellar shell immediately, since the timescale of star formation 
is basically the free-fall time in a high density region, which is considerably 
shorter than the dynamical timescale of the whole cloud. 

The basic equation for the stellar dynamics is given by 
\begin{equation}
	\frac{d^2r_{\rm s}}{dt^2}=-\frac{GM_{\rm tot}(<r_{\rm s})}{r^2_{\rm s}}, 
\end{equation}
where $r_{\rm s}$ and $M_{\rm tot}(<r_{\rm s})$ are the radius of stellar shell 
and the total mass (including baryon and DM components) inside $r_{\rm s}$, 
respectively. 
For stellar shells, we allow them to cross each other. 
We also calculate the mass weighted velocity dispersion, which is represented by
\begin{equation}
	\sigma^2_*= \frac{\sum_n^{N_*} dm_{\rm *,n}v^2_{\rm *,n}}{M_{\rm *,tot}},
\end{equation} 
where $N_*$, $dm_{\rm *,n}$, $v_{\rm *,n}$ and $M_{\rm *,tot}$ are 
the number of stellar shells, mass of $n$-th stellar shell, infall velocity of 
$n$-th shell and the total stellar mass, respectively. 

\subsection{Setup}
The initial density distribution of a cloud consisting 
of gas and DM is the same as that in \cite{Kitayama01}. 
We calculate the evolution of the cloud 
from its linear overdensity stage. 
We concentrate on low-mass clouds 
with the initial baryonic mass of 
$10^5M_{\odot} \la M_{\rm b, in} \la 10^8M_{\odot}$
collapsing at high redshifts $3 \la z_{\rm c} \la 20$. 
Since we define $z_{\rm c}$ as the epoch 
at which a cloud as a whole collapses, 
the collapse redshift of the central high-density regions $z_{\rm c0}$ is earlier than $z_{\rm c}$. 
These two redshifts are related by $1+z_{\rm c0}=2.7(1+z_{\rm c})$. 
When $z\leq z_{\rm UV}$, the cloud is exposed to external UV radiation.
Here, we assume the constant UV intensity specified by $I_{21}$, 
where $I_{21}$ is the intensity at 
the Lyman limit frequency of hydrogen $\nu_{\rm L}$ 
in units of $10^{-21}\rm \;erg\;cm^{-2}s^{-1}Hz^{-1}str^{-1}$. 

The cosmic reionization epoch is inferred to be $z_{\rm r}\approx 11$ 
\citep{Page07,Komatsu09}. 
The reionization of intergalactic matter can start at earlier epochs in
an inhomogeneous fashion \citep{Nakamoto01,Ciardi01}. 
Although the ionization sources are still uncertain, 
the reionization by Pop III stars has been considered as one of
plausible possibilities \citep{Gnedin00,Ciardi01,Cen03,Sokasian04}.
Here, we assume Pop III-type ionization sources.
Previous studies have shown that Pop III stars are as massive as 
$\sim 100M_{\odot}$ \citep[e.g.,][]{ABN00,BCL02,NU01,Yoshida06}. 
Then, the blackbody radiation with effective temperature of
$T_{\rm eff}\simeq 10^5$K is emitted from a Pop III star. 
The number of ionizing photons emitted per second is 
$\dot{N} \sim 10^{50}\rm s^{-1}$ \citep{Schaerer02}. 
Since a Pop III object is as massive as $10^6M_{\odot}$ 
\citep[e.g.,][]{ON99,Yoshida03}, 
almost all ionizing photons emitted from 
a Pop III star can escape from the Pop III object \citep{Kitayama04}.
Qualitatively, the virial radius of the Pop III object is given by 
\begin{equation}
	R_{\rm vir}=1.60\times 10^{2} \left(\frac{M_{\rm f}}{10^6M_{\odot}}\right)^{1/3}
	\left(\frac{20}{1+z_{\rm c}}\right)\rm pc,
\end{equation}
where $M_{\rm f} $ is the total mass of the Pop III object and  $z_{\rm c}$ is its collapse redishift. 
On the other hand, the $\rm Str\ddot omgren$ sphere, 
which is estimated for the cosmic mean density, is extended to
the radius 
\begin{equation}
	R_{\rm s}= 5.45\times 10^3 \left(\frac{\dot{N}}{10^{50}}\right)^{1/3}
 	\left(\frac{20}{1+z}\right)^{2} \rm pc.
\end{equation}
If the radiation flux from a Pop III star is translated into an averaged 
intensity $I_{\nu}$, we have
\begin{equation}
	I_{\nu}=B_{\nu}(T_{\rm eff})\left(\frac{R_*}{r}\right)^2.
\end {equation}
Under the assumption that one Pop III star is born in a halo, 
the UV intensity is evaluated to be from $I_{21}\sim 10^3$ 
at the virial radius of PopIII halo to $I_{21}\sim10^{-3}$ at 
the Str$\ddot {\rm o}$mgren radius.
Hence, we investigate the range of $10^{-3}\leq I_{21} \leq10^3$.
As for UV irradiation epoch, we investigate three cases as 
$z_{\rm UV}$=15, 20, or 25.

\section{Results of Simulations}
\label{results}

\begin{figure}
	\centering
	{\includegraphics[width=7cm]{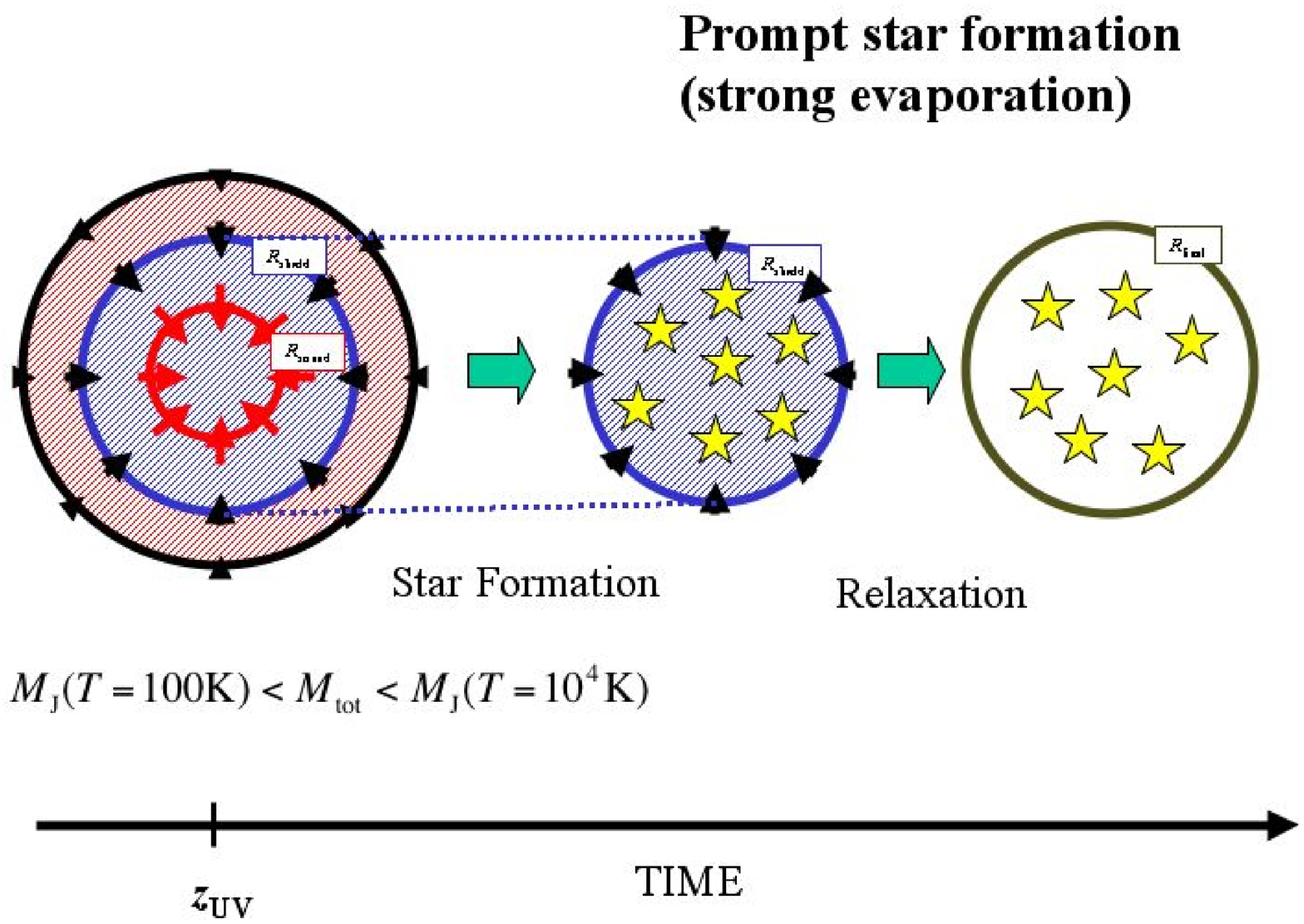}}
	{\includegraphics[width=7cm]{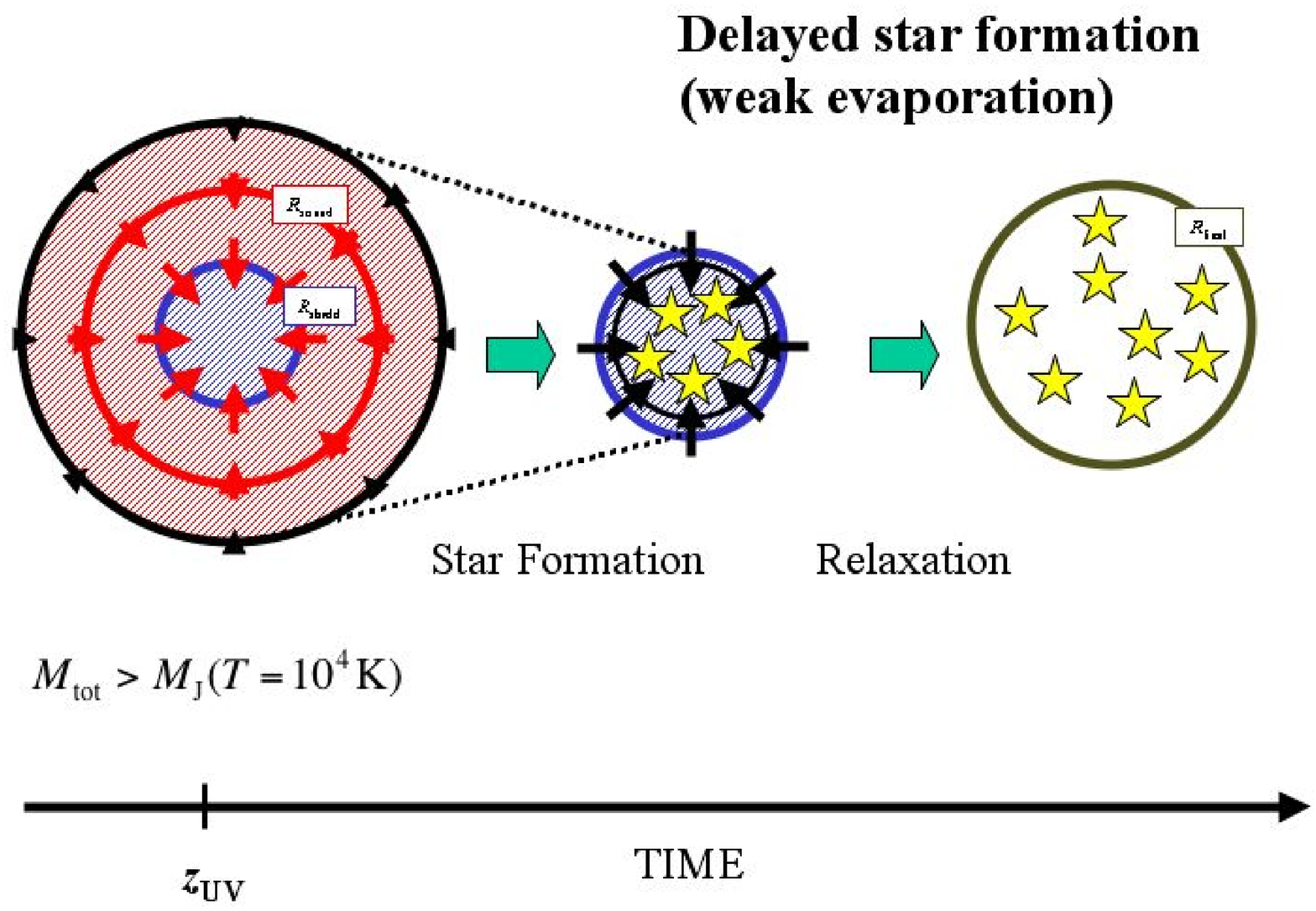}}
	{\includegraphics[width=7cm]{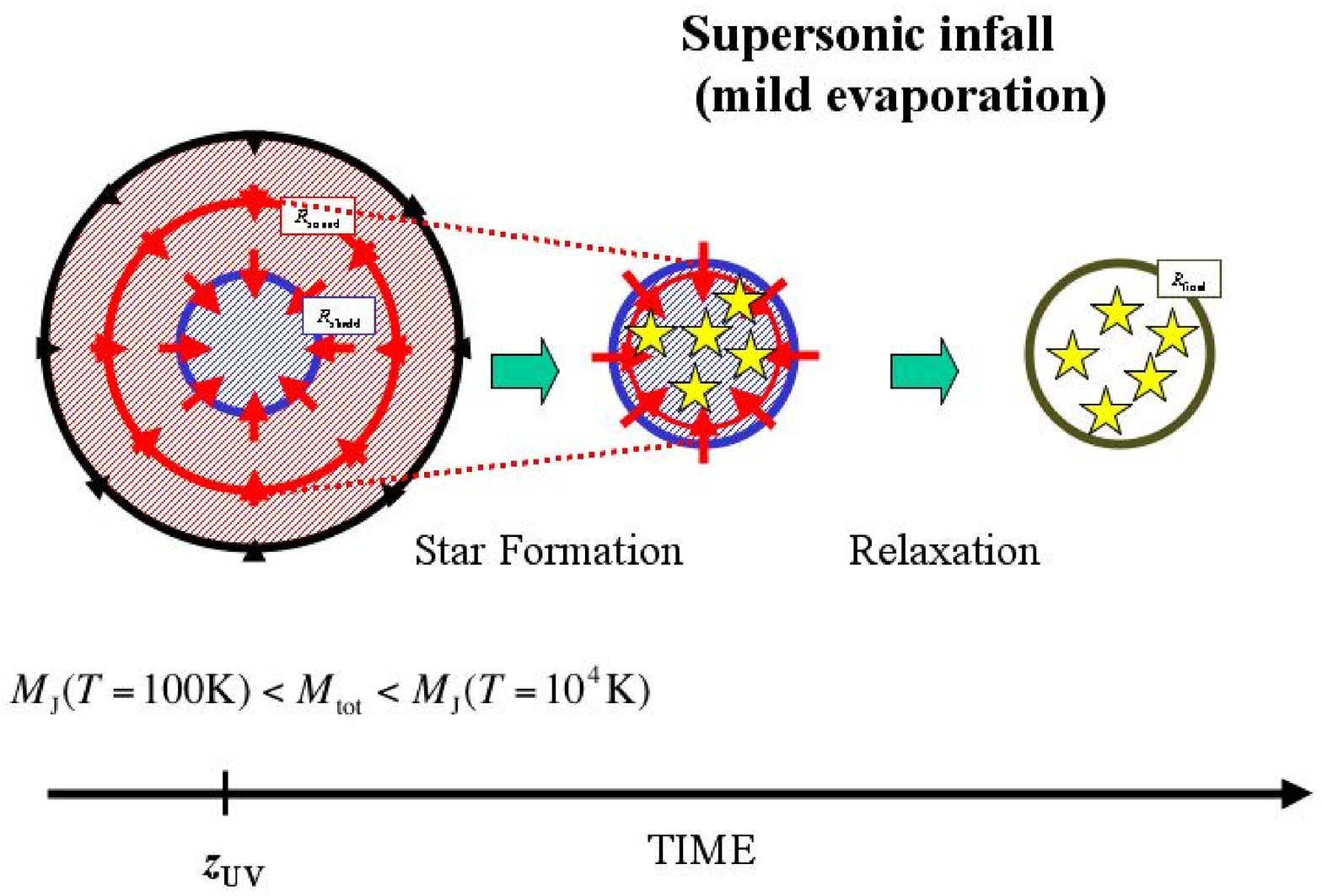}}
	\caption{Schematic views for the {\it prompt star formation} case (top panel), 
	the {\it delayed star formation} case (middle panel), and 
	the {\it supersonic infall} case (bottom panel). 
	In each panel, arrows indicate infall velocity vectors. 
	Blue circles indicate the self-shielding radii $R_{\rm shield}$, inside which 
    the gas is impervious to external UV radiation and
    therefore can cool by H$_2$ cooling to form stars. 
	Red circles indicate the sonic point $R_{\rm sonic}$, 
	where the infall velocity corresponds to the sound speed of $10^4{\rm K}$.} 
	\label{schem}
\end{figure}

\subsection{Hydrodynamics and star formation}
\label{3.1}

If a cloud is not irradiated by UV radiation or 
self-shielded from a UV background, it cools down
to several $100{\rm K}$ by H$_2$ cooling.
Then, a cloud more massive than the Jeans mass
of $100{\rm K}$, $M_J(100{\rm K}) \sim 10^5M_\odot$,
can collapse to form stars. 
When a cloud is photoionized, the gas temperature is raised up to
$\approx 10^4{\rm K}$, and then the Jeans mass increases
to $M_J(10^4{\rm K}) \sim 10^8M_\odot$. 
If the cloud mass is below $M_J(10^4{\rm K})$, the ionized gas cannot
be confined to a DM halo, but evaporated. 
Nevertheless, if the infall velocity exceeds the sound speed of 
$10^4{\rm K}$, the ionized gas can collapse even when
the total mass is below $M_J(10^4{\rm K})$.
Thus, the hydrodynamic evolution of gas cloud and subsequent 
star formation are sensitively dependent on the total mass of cloud, 
the strength of self-shielding, and the infall velocity. 
In the following, we show the typical numerical results, which 
are basically categorized into three types of evolutionary branches. 

\subsubsection{Prompt star formation (strong evaporation)}
\begin{figure}
	\centering
	%\rotatebox[origin=r]{90}
	{\includegraphics[width=7cm]{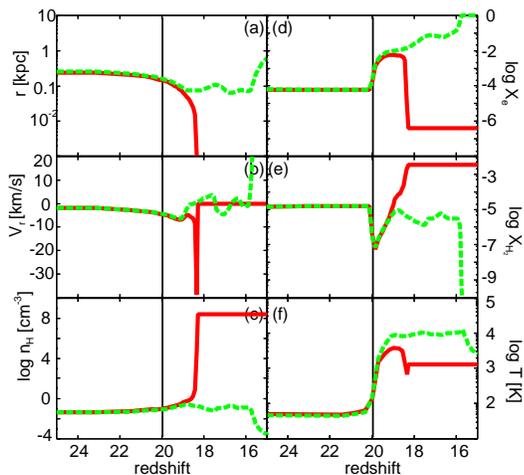}}
	\caption{Cloud evolution in the case of {\it prompt star formation}.
The physical quantities around the self-shielding radius 
are shown as a function of redshift, with
(a) radii, (b) infall velocity, (c) hydrogen number density, 
	(d) electron fraction, (e) H$_2$ fraction, and (f) gas temperature 
	of the cloud. 
	The parameters here are $z_{\rm UV}=20$, $M_{\rm b, in}=10^6M_{\odot}$, 
	$z_{\rm c}=7$, and $I_{21}=1$. 
	In each panel, a thick solid line shows the shell that finally collapses, 
	while a thick dashed line shows the shell that finally evaporates. 
	A thin vertical solid line indicates the epoch of UV irradiation, $z_{\rm UV}$. }
	\label{evo_ps}
\end{figure}

This is the case of the cloud mass between $M_J(100{\rm K})$
and $M_J(10^4{\rm K})$. 
When external UV radiation irradiates the cloud, 
the inner regions at $r<R_{\rm shield}$ are promptly self-shielded
to form stars. But, the outer envelope of cloud is photoionized
and evaporated due to the enhanced thermal pressure, 
because $M<M_J(10^4{\rm K})$.
The schematic view is shown in Fig. \ref{schem} (top panel).
The cloud evolution is shown in Fig. \ref{evo_ps}, where 
the mass shell evolution is shown around the self-shielding radius $R_{\rm shield}$.
After the cloud is irradiated by external UV, a shell inside $R_{\rm shield}$
can collapse, because the self-shielded mass is higher than $M_J(100{\rm K})$.
Eventually, the shell cools down, forming stars promptly. 
In this case, the star formation criteria is satisfied from inside out to
$R_{\rm shield}$. 
The duration of star formation is roughly 10-100Myr.  
The duration is longer for higher mass clouds, 
since a larger part of massive clouds can be shielded. 
On the other hand, the regions outside $R_{\rm shield}$ are ionized and
photoheated. Then, shells there evaporate shortly.

As a result, the mass-to-light ratios are expected to be higher owing to
the mass loss by the photoevaporation.
Such behaviors can be basically understood as the formation mechanism 
of dwarf galaxies in a UV background that is previously studied in detail 
by several authors \citep[e.g.,][]{Kitayama01,SU04,RGS08}.

\subsubsection{Delayed star formation (weak evaporation)}
\begin{figure}
	\centering
	%\rotatebox[origin=r]{90}
	{\includegraphics[width=7cm]{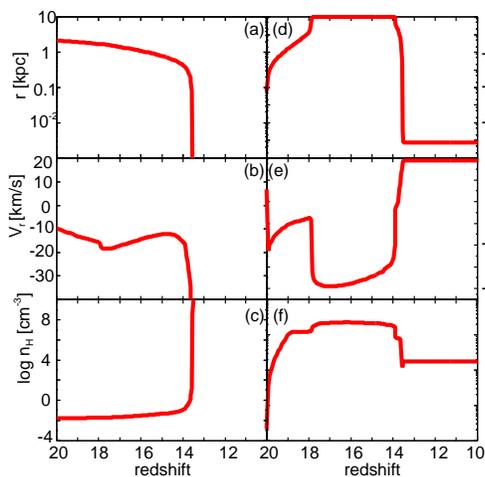}}
	\caption{Cloud evolution in the case of {\it delayed star formation}.
    The physical quantities same as those in Fig. \ref{evo_ps} are shown.
    The parameters here are 
	$z_{\rm UV}=20$, $M_{\rm b, in}=10^8M_{\odot}$, 
	$z_{\rm c}=7$, and $I_{21}=100$. 
	The time evolution of the outermost shell that finally collapses is shown.}
	\label{evo_ds}
\end{figure}

This is the case of cloud mass slightly higher than $M_J(10^4{\rm K})$. 
When UV intensity is very weak, the almost all regions of cloud are self-shielded
and collapse in a similar way to the UV-free case. 
However, if external UV is relatively strong, the bulk of cloud is photoionized.
No star forms in the ionized regions, since no H$_2$ cooling works there.
Nonetheless, most of ionized regions do not evaporate but collapse,
because the cloud is more massive than $M_J(10^4{\rm K})$ and therefore
the self-gravity overwhelms the pressure of the ionized gas. As a result,
in the course of cloud contraction, the ionized regions are self-shielded 
due to the increase of density. 
Then, stars form there in a delayed fashion. 
In this case, the star formation can continue for $>100{\rm Myr}$
because of the delayed self-shielding. 
The schematic view is shown in Fig. \ref{schem} (middle panel).
The cloud evolution in a strong UV background is shown in Fig. \ref{evo_ds}, 
where the outermost shell that finally collapses is shown.
The shell is ionized after $z_{\rm UV}=20$, but it collapses and is self-shielded 
around $z=14$, forming a cooled shell.
As a result of such delayed self-shielding, a formed star cluster can become 
more compact than a star cluster formed in no UV background, and
therefore the stellar velocity dispersion of cluster is increased.  
It is noted that the effect of delayed star formation becomes
conspicuous for the mass slightly above $M_J(10^4{\rm K})$,
since a much higher mass cloud is impervious to photoionization
except for surface thin layer.

\subsubsection{Supersonic infall (mild evaporation)}
\begin{figure}
	\centering
	%\rotatebox[origin=r]{90}
	{\includegraphics[width=7cm]{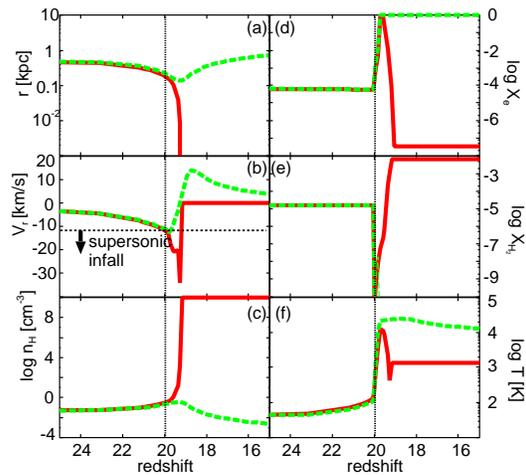}}
	\caption{Cloud evolution in the case of {\it supersonic infall}.
	Same as Fig. \ref{evo_ps}, except for $M_{\rm b, in}=10^7M_{\odot}$
	and $I_{21}=100$. 
	In panel (b), the sound speed of $10^4{\rm K}$ is shown by the horizontal 
	dotted line. The time evolution of shells around the sonic point is shown.}
	\label{evo_ss}
\end{figure}

This is a novel branch that is found in the present simulations. 
The cloud mass is between $M_J(100{\rm K})$ and $M_J(10^4{\rm K})$,
similar to the case of prompt star formation, but 
the infall velocity of the cloud can exceed the sound speed of $10^4$K 
owing to radiative cooling as well as a potential of DM halo. 
When the cloud is exposed to strong UV radiation, ionizing photons
permeate the cloud deeply, and a large part of cloud is photoionized.
However, shells collapsing with supersonic infall velocities cannot
be stopped by photoheating. Therefore, they keep contracting,
regardless of photoionization. Such shells are eventually 
self-shielded, when the density
increases sufficiently due to the contraction, and then forms a compact
star cluster.
The schematic view of the supersonic infall case is shown 
in Fig. \ref{schem} (bottom panel).
The evolution of shells around the sonic point
is shown in Fig. \ref{evo_ss}.
A dotted line shows an evaporating shell, while a solid line 
shows a collapsing shell with supersonic infall velocity.
It is clear that the supersonic infall shell is photoionized
once (panel b) , and eventually self-shielded 
from both ionizing photons (panel d) and 
$\rm H_2$-dissociating photons (panel e), and cools below 
$10^4\rm K$ (panel f) by $\rm H_2$ cooling. 
Hence, the gas inside the sonic point $R_{\rm sonic}$ results
in a compact star cluster. 

As a result, the strong contraction before star formation 
enhances the stellar velocity dispersion to a large degree. 
As a matter of importance, in this {\it supersonic infall} case,
the duration of star formation
becomes less than 10Myr, in contrast to the other two cases. 
This means almost coeval star formation, which is consistent with 
the fact that GCs commonly have a single stellar population. 
The {\it supersonic infall} branch appears
if the sonic point is larger than the self-shielding radius.

\begin{figure}
	\centering
	{\includegraphics[width=7cm]{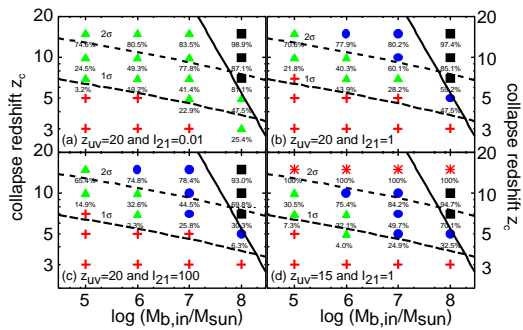}}
	\caption{Dependence of cloud evolution on the cloud mass, collapse epoch, 
	UV intensity, and UV irradiation epoch. 
	The collapse redshift $z_{\rm c}$ and the type of evolution are 
	shown against the initial baryonic mass of cloud, $M_{\rm b,in}$. 
	Panels (a), (b) and (c) show the results with $z_{\rm UV}=20$,
  	  but for different UV intensity as $I_{21}=0.01$, 
	$I_{21}=1$, and $I_{21}=100$, respectively. 
	Panel (d) is the result of $z_{\rm UV}=15$ and $I_{21}=1$.
	Triangles, squares, and circles indicate
	the cases of {\it prompt star formation}, 
	{\it delayed star formation}, and {\it supersonic infall}, respectively. 
	Pluses represent complete evaporation, and asterisks correspond to evolution in no UV. 
	The percentages attached to the symbols represent the fraction of 
    	the final stellar mass $M_*$ to the initial baryonic mass, $M_*/M_{\rm b,in}$.
	In each panel, a long-dashed and a short-dashed line correspond to
	1$\sigma$ and 2$\sigma$ CDM fluctuations. A solid line represents 
	the Jeans mass of ionized gas $M_{\rm J}(10^4{\rm K})$.} 
	\label{numsum}
\end{figure}

\subsection{Parameter Dependence}

The dependence of cloud evolution on the mass, collapse epoch, UV intensity, 
and UV irradiation epoch are summarized in Fig. \ref{numsum}. 
In each panel, the resultant types of evolution are shown
in the diagram of the collapse redshift against the initial baryonic
mass of cloud. Also, $1\sigma$ and $2\sigma$ CDM density
fluctuation spectra are shown. 
The UV intensity is changed with $I_{21}=10^{-2},1$ or $10^2$. 
In the panels (a), (b) and (c),  the UV irradiation epoch is assumed
to be $z_{\rm UV}=20$, while $z_{\rm UV}=15$ in the panel (d).  
In the simulations with $z_{\rm UV}=15$, the objects collapsed 
before $z=15$ correspond to the UV-free case. 
Interestingly, the boundary between the complete evaporation and star cluster
formation roughly corresponds to $1\sigma$ CDM fluctuation. 

In each panel, a thin solid line indicates the Jeans mass of 
ionized gas $M_{\rm J}(10^4{\rm K})$, which is given by
\begin{equation}
	M_{\rm J}(10^4{\rm K})=1.1\times10^9 M_{\odot}(1+z_{\rm c})^{-3/2}
\left(\frac{\Omega_{\rm b}} {0.05} \right) 
\left(\frac{\Omega_{\rm M}} {0.3} \right)^{-1} . 
\end{equation}
Above this line, the cloud evolution results in the {\it delayed star formation}
as seen in Fig. \ref{numsum}.
Below the Jeans mass $M_{\rm J}(10^4{\rm K})$, 
clouds with more massive and higher collapse epochs 
lead to the star cluster formation by the {\it supersonic infall} .
This tendency can be understood by following two reasons:  
(i) The self-shielding radius is weakly dependent on the cloud mass 
($R_{\rm shield} \propto M^{2/3}$) shown by \citet{TU98}, 
while the radius where the infall velocity exceeding the sound speed 
is roughly proportional to the cloud mass. Hence, a higher mass cloud 
has larger ionized, supersonic regions.
(ii) The fraction of supersonic infall regions becomes 
larger according as the cloud contracts. Thus, earlier collapsing clouds
tends to form most stars through supersonic infall. 

In $z_{\rm UV}=15$ simulations, 
the trend is the same as the simulations with $z_{\rm UV}=20$,
as long as the collapse epoch is later than $z_{\rm UV}=15$.
On the other hand, in the case of weak UV with $I_{21}=10^{-2}$
(panel a), the {\it supersonic infall} case does not appear,
because the self-shielding radius is always larger 
than the sonic point. 

\subsection{Stellar Dynamics}
\begin{figure}
	\centering
	{\includegraphics[width=7cm]{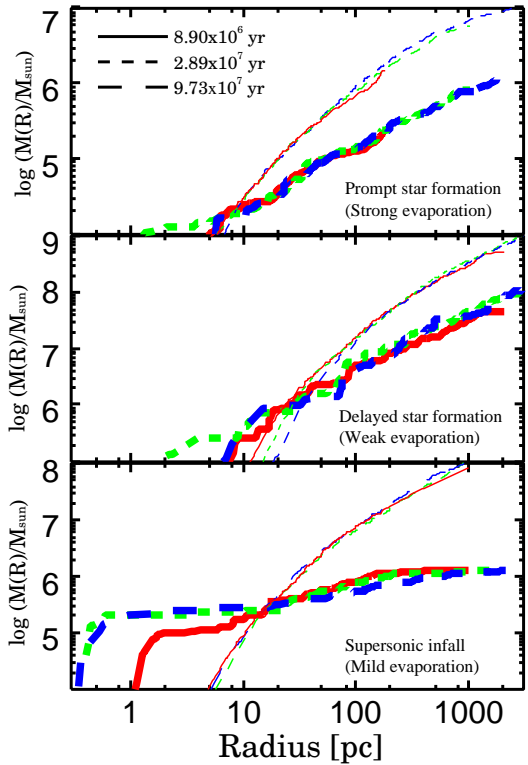}}
	\caption{Mass distributions for typical three cases:
	{\it prompt star formation} case (top panel), 
	{\it delayed star formation} case (middle panel), and 
	{\it supersonic infall} case (bottom panel). 
	In each panel, time variations of mass distributions are
	shown by $M(R)$ in units of the solar mass, where $M(R)$
	is the cumulative mass in the regions of $r \leq R$.
    	Thick lines show stellar components, while thin lines do DM components.
	Solid, short-dashed, and long-dashed lines are respectively corresponding to
	the distributions at $8.9\times 10^6\rm yr$, $2.89\times 10^7\rm yr$ 
    and $9.73\times 10^7\rm yr$ after $z_{\rm UV}$.}
	\label{profiles} 
\end{figure}

Here, we pursue the subsequent stellar dynamics of star clusters
formed through three mechanisms shown above.
In Fig. \ref{profiles}, 
the time sequence of mass distributions is shown for {\it prompt star formation},
{\it delayed star formation}, and {\it supersonic infall} cases.
Obviously, the mass distribution of a star cluster formed 
through the {\it supersonic infall} is different from the other two cases. 
In the {\it supersonic infall} case, the stellar component is predominant 
in the inner several $10\rm pc$ regions. 
On the other hand, in the other two cases, the system is 
dominated by the DM component 
in almost all regions. These results are understood as follows: 
In the {\it supersonic infall} case, 
the star formation is delayed by the external UV radiation, 
until the cloud contraction eventually causes the self-shielding.
Then, the strong energy dissipation occurs in a compact region, 
leading to the formation of a dense star cluster. 
As a result, a compact star-dominant system forms. 
On the other hand, in the {\it prompt star formation} case, 
stars are born at earlier dynamical phase of initially self-shielded 
regions. Consequently, the energy dissipation is not so strong 
and therefore a more diffuse star cluster forms. 
In the {\it delayed star formation} case, the cloud is self-shielded
after it contracts to some degree. Hence, the dark matter contribution
becomes smaller compared to the {\it prompt star formation} case.
But, it does not grow into a totally star-dominant system, 
since the cloud is self-shielded before it becomes baryon-dominated. 
Thus, we conclude that the {\it supersonic infall} is only the branch
that produces star-dominant compact clusters. 

\section{Comparison with Observations}
\begin{figure}
	\centering
	{\includegraphics[width=7cm]{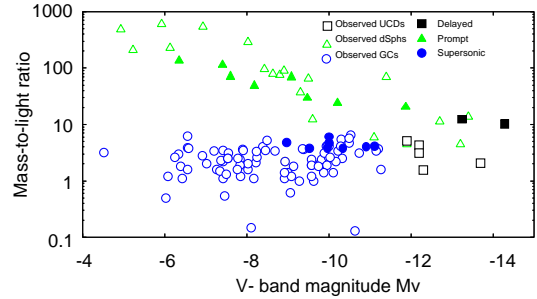}}
	\caption{Mass-to-light ratios as a function of absolute V-band magnitude $M_{\rm V}$. 
	Large filled symbols are simulations, while small open symbols are observations.
    Filled triangles, filled squares, and filled circles indicate
	the objects formed through {\it prompt star formation}, {\it delayed star formation}, 
	and {\it supersonic infall}, respectively. 
	On the other hand, open triangles, open squares, and open circles 
	indicate observed dwarf spheroidals (dSphs), ultra-compact dwarfs (UCDs),
    globular clusters (GCs), respectively. 
	The observational data of GCs are taken from the data of Milky Way GCs \citep{PM93}, 
	M31 GCs \citep{Fischer93,Dubath96,Djo97}, 
	LMC GCs \citep{DG97,Fischer93},  
	SMC GCs \citep{Dubath92}, and NGC5218 GCs \citep{MH04}. 
	The data of dSphs are taken from \citet{Mateo98}, \citet{Matin06},
	\citet{SG07}, \citet{MCI06}, and \citet{Majewski07}. 
	The UCD data are taken from \citet{Drinkwater03}.
	} 
	\label{mlvsmv}
\end{figure}
\begin{figure}
	\centering
	{\includegraphics[width=7cm]{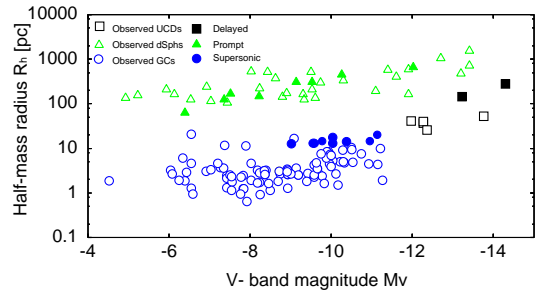}}
	\caption{Half-mass radii $R_{\rm h}$ as a function of absolute V-band 
	magnitude $M_{\rm V}$. 
    	The symbols are the same sense as Fig. \ref{mlvsmv}.} 
	\label{rhvsmv}
\end{figure}

\subsection{Mass-to-light ratios and half-mass radii}

As shown in \S3, the final mass-to-light ratios 
depend on the formation processes. 
Here, we derive the mass-to-light ratios of simulated objects.
In this section, we use simulation results with $z_{\rm UV}=20$, 
$I_{21}=10^{-2}, 1$, or $10^2$, $M_{\rm b,in}=10^5M_{\odot}-10^8M_{\odot}$, 
and $z_{\rm c}=5$ or 7, which corresponds to 
the central collapse redshift of $z_{\rm c0}\approx 15.2$ or $20.6$, respectively.  
For each simulated object, the dynamical mass is evaluated by 
the total mass ($M_{\rm dyn}=M_*+M_{\rm DM}$) 
inside the half-mass radius of the stellar component.
As for stellar components, we assume $M_*/L_{\rm V}=2$, 
where $M_*$ is the total stellar mass and $L_{\rm V}$ is 
the V-band luminosity \citep{PM93}. 
This is a typical value when stars produced by initial starbursts evolve 
for 10 Gyr.
In Fig. \ref{mlvsmv}, the resultant mass-to-light ratios $M_{\rm dyn}/L_{\rm V}$ 
of simulated objects 
are shown as a function of absolute V-band magnitude $M_{\rm V}$, 
and they are compared to the observations of globular clusters (GCs),
dwarf spheroidals (dSphs), and ultra-compact dwarfs (UCDs). 

As seen clearly, the models of
{\it prompt star formation}, {\it delayed star formation}, and {\it supersonic infall}
are distinctively separated in this diagram. 
In particular,  the model of {\it supersonic infall} matches fairly well with bright GCs. 
It is worth noting that even if we change $M_*/L_{\rm V}$ by a factor of 2, 
it does not alter basic results significantly. 
The models of {\it prompt star formation} and {\it delayed star formation} 
seem to match dSphs and UCDs, respectively in this diagram. 
However, these two models can be affected by internal feedbacks. 
The effects by the internal feedback are discussed meticulously in the next section. 
Anyhow, if the final $M_*/L_{\rm V}$ for stellar components is in the
range of $1 \la M_*/L_{\rm V} \la 4$, three models are distinctive
in this diagram. The mass loss by internal feedbacks 
in {\it prompt star formation} and {\it delayed star formation} 
would result in an up-shift in Fig. \ref{mlvsmv}, 
leading to a further deviation from GCs. 

Also, in Fig. \ref{rhvsmv}, the resultant half-mass radii $R_{\rm h}$ 
of simulated objects are shown as a function of absolute V-band 
magnitude $M_{\rm V}$, and compared the observations.
Again, the model of {\it supersonic infall} is well concordant with observed bright GCs, 
whereas the other two models are not. 
Here, a remarkable property for the {\it supersonic infall} model 
is that the half-mass radius is almost independent of the mass. 
This is due to the fact that a higher mass cloud has larger ionized, 
supersonic region, and eventually the energy dissipation occurs more strongly. 
It is worth noting that half-mass radii in {\it prompt star formation} case are 
about tens times larger than those in {\it supersonic infall} case 
for the same luminosity range. 
These results imply that UV background radiation works significantly 
to determine the properties of subgalactic objects. 
The final mass-to-light ratios and half-mass radii are both responsible
for the velocity dispersions of objects. 

\subsection{Velocity dispersions}
\begin{figure}
	\centering
	{\includegraphics[width=8cm]{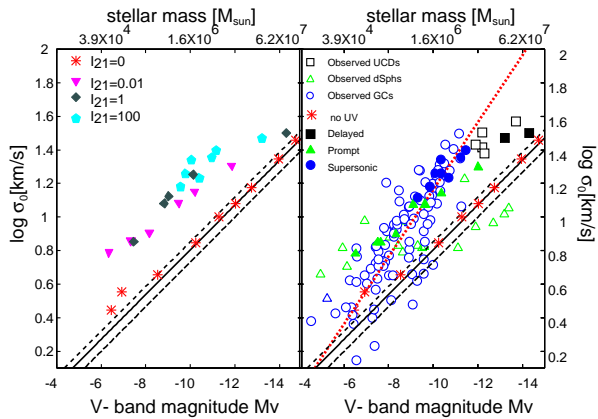}}
	\caption{Velocity dispersions $\sigma_*$ as a function of 
	absolute V-band magnitude $M_{\rm V}$. 
	In the left panel, the dependence on UV intensity is shown.
    Asterisks, reverse triangles, diamonds, and 
	pentagons represent the results for $I_{21}=0$, $I_{21}=0.01$, $I_{21}=1$, and 
	$I_{21}=100$, respectively. 
	The virial relation expected in the UV-free case are also shown
	by a long-dashed, a solid, and 
	a short-dashed line, respectively for the collapse epoch of
    $z_{\rm c}=$5, 7, and 10. 
    In the right panel, the simulation results are compared with 
    the observational data, where the meanings of symbols are 
    the same as Fig. \ref{rhvsmv}. The best fit relation for the 
	observed GCs is shown by a dotted line. 
	} 
	\label{vdvsmv}
\end{figure}

In Fig. \ref{vdvsmv}, the stellar velocity dispersion $\sigma_*$ of 
simulated objects are shown 
as a function of absolute V-band magnitude $M_{\rm V}$. 
Here, we also show the results with $I_{21}=0$ (UV-free case), 
assuming $M_{*}/L_{\rm V}=2$. 
In the left panel, the dependence on UV background intensity is shown 
for $I_{21}=0$, $I_{21}=0.01$, $I_{21}=1$, and $I_{21}=100$.
Also, the relation ($L \propto \sigma^3_*$) predicted 
from the virial theorem in the UV-free case is shown
for the collapse redshifts of $z_{\rm c}=$5, 7, and 10. 
The objects without UV effects well follow this prediction. 
In the presence of the UV background radiation, 
systematically larger velocity dispersions are predicted for a given 
V-band magnitude. 
%As seen clearly in this figure, the UV background 
%significantly raises the velocity dispersions. 

In the right panel, the simulation results are
compared with observational data. 
The objects formed through {\it supersonic infall} are distributed near 
the best fit relation for the observed GCs shown by a dotted line. 
It should be noted that only the {\it supersonic infall} case can account
for GCs with high velocity dispersions $\ga 10$ km s$^{-1}$. 
For the {\it supersonic infall} case, higher mass clouds result in
the stronger energy dissipation. 
Consequently, the relation between $\sigma_*$ and $M_{\rm V}$ 
becomes steeper than that the UV-free case ($\sigma_* \propto L^{1/3}$). 
The correlation for the {\it supersonic infall} case is roughly 
given by $\sigma_* \propto L^{1/2}$, which is consistent with the observed GCs 
\citep[e.g.,][]{MvdM05}. 
Since velocity dispersions are estimated as $\sigma_* \propto \sqrt{GM/R}$,
this relation implies that the radii of objects is almost regardless of the mass.
This is consistent with the resultant half-mass radii $R_{\rm h}$ 
shown in Fig. \ref{rhvsmv}.  

The velocity dispersions for the {\it delayed star formation} case 
are systematically larger than the prediction by the simple virial theorem, 
since the energy dissipation is stronger than that in the UV-free case.   
In this diagram,  the velocity dispersions for {\it supersonic infalling} case 
and {\it prompt star formation} case are degenerated at 
$-8\ge M_{\rm V} \ge -10$. 
But, the origin is different, since the distributions in mass-to-light ratios and 
half-mass radii are well separated as shown above. 
Since the outer regions of cloud evaporates in
the {\it prompt star formation} case, 
the velocity dispersion is determined basically by the DM component.

In Figs. \ref{mlvsmv}, \ref{rhvsmv}, and \ref{vdvsmv}, 
the {\it supersonic infall} does not reproduce 
faint GCs in the range of $-6>M_{\rm V}>-9$. 
If we subdivide the parameter space near the low-mass boundary 
between the models of 
{\it supersonic infall} and {\it prompt star formation},
compact star clusters like faint GCs possibly appear. 
Another possibility is the mass loss through tidal stripping by a host galaxy,
which results in low-mass GCs.  

\section{Discussion}

\subsection{Specific frequencies}

It is known that early-type galaxies have higher specific frequency of GCs than
late-type galaxies  \citep{Harris91}. 
Moreover, the specific frequency depends on the luminosity of host galaxy 
\citep{Forbes05,Bekki06}. 
The specific frequency $S_{\rm N}$ is defined as the GC population normalized to 
$M_{\rm V,host}=15$ as 
\begin{equation}
	S_{\rm N} \equiv N_{\rm t} \times 10^{0.4(M_{\rm V,host}+15)} , 
\end{equation}
where $N_{\rm t}$ is the total number of GCs in a host galaxy and 
$M_{\rm V,host}$ is the V-band magnitude of host galaxy. 
This tendency could be qualitatively understood by 
combining the present results with the galaxy formation
theory in UV background radiation, which is studied by \citet{SU00}. 
\citet{SU00} have shown that early-type galaxies form more preferentially 
at high-$\sigma$ peak regions of CDM fluctuations,
since the self-shielding works effectively to allow the high efficiency of star formation 
and therefore the galaxy formation proceeds in a dissipationless fashion. 
The present simulations have shown that GCs can also form preferentially
from high-$\sigma$ fluctuations ($> 2\sigma$)  as shown in Fig. \ref{numsum}. 
Therefore, it is expected that the number of GCs around 
an early-type galaxy tends to be larger than that in a late-type galaxy. 

\citet{Moore06} have explored the spatial distribution of subhalos in host galaxies, 
by high-resolution $N$-body simulations. 
They have found that the spatial distributions of subhaloes originating in 
high-$\sigma$ peaks ($>2.5\sigma$) are similar to 
those of old metal-poor GCs in the Milky way galaxy. 
Thus, it is likely that objects by the {\it supersonic infall} can correspond 
to the metal-poor GCs in host galaxies. 

\citet{Saitoh06} have carried out hydrodynamic simulations on the galaxy formation 
with high spatial resolution. 
It is found that numerous clumps of globular cluster mass scale form in a host galaxy.
The specific frequency $S_{\rm N}$ can be roughly estimated to be the $S_{\rm N}\sim 20$,
assuming the mass-to-light ratio of $M/L_{\rm V}=5$. 
On the other hand, the specific frequency in spiral galaxies is observed 
to be $S_{\rm N}\simeq 1$ \citep{Harris91}.
This implies that not all the small mass clumps do not evolve into GCs. 
Here, we have shown that only the {\it supersonic infall} branch can lead to
the formation of GCs. Hence, a portion of small mass clumps evolve into
GCs. To make quantitative argument, it is necessary to perform high 
resolution simulations on the galaxy formation including the radiative transfer
effect of UV background radiation. This would be a future challenge. 

\subsection{Sites of GC formation}

As shown in Fig. \ref{numsum}, the {\it supersonic infall} case 
does not appear in the simulations with $I_{21}\leq 0.01$. 
If there is no or very weak UV radiation fields, 
only low density star clusters form (top panel in Fig. \ref{profiles}). 
To form compact star clusters like GCs, strong UV radiation, 
roughly $I_{21} > 1$ is required. 
Massive Population III stars are one possibility to produce strong UV intensity 
in high redshift epochs. But, the strong UV radiation is relatively
localized in HII region around a Population III halo \citep{SU06, HUS09}. 
Another possibility is black hole accretion that could be a more powerful source
\citep{RO04, SK00}.
\citet{SK00} simply estimated the UV intensity around AGN of 
$10^{44}$erg s$^{-1}$ as 
\begin{equation}
	I_{\nu_{\rm L}}\sim 1\times \left(\frac {100{\rm kpc}}{R_{\rm g}}\right)^2
	\times 10^{-21} {\rm erg\: s^{-1}cm^{-2}str^{-1}},
\end{equation}  
where $R_{\rm g}$ is the distance from the black hole.
That is to say, $I_{21}\sim 1$ is achieved at $100\rm kpc$ around an AGN. 
This can allow the GC formation in a fairly wide area around the host galaxy. 

\subsection{Tidal stripping}

It is expected that the evolution of star clusters can 
be affected by the tidal force by its host galaxy 
(i.e. Mashchenko \& Sills 2005). 
The importance of the tidal interaction is estimated as follows.
The balance between the self-gravity and the tidal force is given by
\begin{equation}
	\frac{Gm^2(<r_{\rm t})}{r^2_{\rm t}}\approx
	\frac{2GM_{\rm host}m(<r_{\rm t})r_{\rm t}}{R^3_{\rm c}},
\end{equation}
where $r_{\rm t}, m(<r_{\rm t}), M_{\rm host}$ and $R_{\rm c}$ are
the tidal radius, the cluster mass inside $r_{\rm t}$, the host galaxy mass and 
the position of star cluster from the centre of host galaxy, respectively. 
GCs are likely to form in small proto-galaxies in the 
early universe \citep[e.g.,][] {BC02,MS05}.  
Hence, we assume that 
$M_{\rm host}=10^{9}M_{\odot}$ and $R_{\rm c}=0.3-1\rm kpc$. 
If we use the distributions obtained by the present simulations, 
the tidal radii $r_{\rm t}$ are evaluated as 10pc - 50pc for 
the {\it supersonic infall} case (bottom panel in Fig. \ref{profiles}). 
This value is larger than the half-mass radii of GCs 
shown in Fig. \ref{rhvsmv}.
This implies that the estimation of mass-to-light ratios 
or velocity dispersions for the {\it supersonic infall} case
is not affected, even if 
the outer diffuse DM halo is stripped away by the tidal force.
Unfortunately, it is difficult to urgue precisely 
the evolution of star clusters in 
the tidal fields by one-dimensional simulations. 
Hence, we need to simulate the three-dimensional dynamical evolution
by means of $N$-body method. 
We are going to present such $N$-body simulations in a forthcoming paper. 
%For the objects formed through {\it delayed star formation}, 
%the tidal stripping process can make the mass-to-light ratios small. 
%This will bring our predictions in Fig. \ref{mlvsmv} even closer to the observed UCDs. 
%For the {\it prompt star formation} case, 
%on the other hand, the simulated objects might be easily 
%destroyed by the tidal interaction with its host galaxy, 
%since their mass distributions are more diffuse than those 
%of {\it supersonic infall} case. 

\subsection{Internal feedbacks}

Our calculations do not include the internal feedback, 
i.e., UV radiation from internal massive stars and supernova (SN) explosions, 
which becomes important in the case of successive star formation. 
%Hence, not only external but also internal feedbacks 
%should be taken into account in the case of successive star formation. 
As mentioned above, the duration of star formation in {\it supersonic infall} 
case is typically several Myrs. The binding energy of a star cluster
produced by {\it supersonic infall} is in the order of $10^{51}{\rm erg}$, 
which corresponds to the typical SN energy. 
Hence, only one massive SN can sweep away the interstellar gas.
Therefore, the {\it supersonic infall} results in a single stellar population,
which is consistent with observed GCs. 
It is also important to mention that such a short duration of star formation 
may also be responsible for the chemical homogeneity  observed in individual GCs. 

On the other hand, much larger stellar metallicity dispersions found 
in dwarf galaxies may be due to successive star formation. 
In the cases of {\it prompt star formation} and 
{\it delayed star formation}, the duration of star formation
is $10-1000$Myr, and the binding energy roughly ranges from 
$10^{51}{\rm erg}$ to $10^{54}{\rm erg}$.
The importance of internal feedbacks is likely to depend on
the mass scale.
In the case of $M_{\rm b, in}<10^7M_{\odot}$, the duration of 
star formation is shorter (several $10$Myr) as shown in \S\ref{3.1}, 
since the star formation is quickly quenched by the external UV radiation.
In addition, SN explosions as well as stellar UV radiation would 
significantly suppress subsequent star formation \citep{Kitayama04,KY05}. 
Interestingly, a recently discovered "faint" dSph Ursa Major I 
shows a single old stellar population \citep{Okamoto08}. 
The low-luminous {\it prompt star formation} model might 
correspond to such "faint" dSphs. 
On the other hand, in the case of $M_{\rm b, in}>10^7M_{\odot}$,
successive star formation is possible even if the cloud is irradiated 
by the external UV radiation. 
Moreover, repeated SN explosions, which might trigger the star formation, 
are possible owing to the fact that the binding energy is larger than $\sim 10^{51}{\rm erg}$.   
Hence, in this case, the present simulations 
should not be directly compared with observations.
The final states would be determined by 
internal feedbacks as shown in \cite{DLM03}. 
They have studied the effect of the internal feedbacks
by UV and SN explosions of formed stars as well as the external UV feedback 
on the formation of dSphs, by spherical symmetric simulations. 
They have shown that the star formation in low-mass dSphs 
is severely suppressed by the external UV radiation, 
whereas the star formation in high-mass dSphs 
is self-regulated by the internal feedback processes, 
and the final equilibria are determined solely by the internal feedback. 
Interestingly, their simulations have succeeded in reproducing multiple stellar 
population, which is observed in some dSphs in the Local Group 
\citep[e.g.,][]{Grebel97,IA02,Grebel04,Tolstoy04}. 

This argument can be also applied in {\it delayed star formation} case.
Although the final state is likely to be determined by the internal feedbacks, 
the external UV works to delay the onset of star formation.  
Owing to such an effect, the energy dissipation becomes slightly strong, 
which leads to the enhancement of velocity dispersion 
at a central region (middle panel of Fig. \ref{profiles}). 
Observed UCDs \citep{Drinkwater03} might be explained by this process
(see Fig. \ref{vdvsmv}).

While the important processes of internal feedbacks has been 
studied by \cite{DLM03} in spherical symmetry, there are also
three-dimensional effects. 
\cite{SU04} performed three-dimensional radiation hydrodynamic
simulations with solving radiative transfer to investigate
the shielding effect of local density peaks within a UV background. 
They have shown the local shielding allows the long duration of star formation
even after the reionization. 
In addition, interstellar medium is compressed intricately by 
shocks produced by multiple SN explosions, and therefore the star formation would be 
triggered by such explosions \citep{MFM02}. 
Moreover, we mention that the dynamics of stars would be more complex, 
owing to three-dimensional effects \citep{ALP88,ML96}. 
These three-dimensional effects can play important roles
for the star formation history and the dynamics of stars. 
Thus, we plan to carry out three-dimensional RHD simulations 
including the external radiation as well as the internal feedback 
to explore the star formation history comprehensively. 

\section{Summary}

We have carried out the radiation hydrodynamic simulations to 
explore the possibility that the formation of GCs is induced by
external UV radiation fields. As a result, 
we have found that the {\it supersonic infall}
enables a low-mass gas cloud to form a compact star
cluster in the external UV radiation.
A gas cloud with its infall velocity exceeding 
the sound speed can keep contracting,  
even if the gas cloud is fully ionized without the self-shielding 
from external UV radiation.
The contracting cloud is shielded from a UV background when
a compact cloud core forms, and cools by $\rm H_{2}$ cooling.
Consequently, a compact star cluster forms in a diffuse DM halo. 
We have also calculated the dynamical evolution of stars. 
It is found that resultant mass-to-light ratio, half-mass radius, and velocity dispersion 
of simulated star clusters match well with those of observed GCs.
Therefore, the {\it supersonic infall} in a UV background 
is a promising mechanism to form GCs.

\section*{Acknowledgments}
We would like to thank N. Arimoto, H. Hirashita, D. N. C. Lin, T. Nakamoto, M. Ricotti, 
and K. Yoshikawa for valuable comments that
considerably improved the manuscript.
Numerical simulations have been performed with computational facilities 
at Centre for Computational Sciences in University of Tsukuba. 
This work was supported in part by the FIRST project based on
Grants-in-Aid for Specially Promoted Research by 
MEXT (16002003), and Grant-in-Aid for Scientific 
Research (S) by JSPS  (20224002). 
Also, it was supported in part by Grant-in-Aid for 
young Scientists (B) by MEXT (18740112, 21740139).

\label{lastpage}

\end{document}